\documentclass[superscriptaddress,notitlepage,amsmath,amssymb]{revtex4-1}

\usepackage{graphicx}
\usepackage{chemarr}
\usepackage{times,longtable}
\usepackage{hyperref}
\usepackage{color}
\usepackage{afterpage}
\usepackage{soul,comment}
\usepackage[usenames,dvipsnames]{xcolor}
\usepackage{colortbl}
\usepackage{pbox}
\usepackage{capt-of}

\graphicspath{{large/}}

\bibpunct{[}{]}{,}{n}{}{,}

\usepackage{chemarr}
\usepackage{epsf,mathtools}
\usepackage{graphicx}
\usepackage{dcolumn}
\usepackage{bm}

\begin{document}

\title{Statistical-physics approaches to RNA molecules, families and networks}




\author{Simona Cocco}
\affiliation{Laboratoire de Physique de l'Ecole Normale Superieure, PSL and CNRS, Paris, France}

\author{Andrea De Martino}
\affiliation{DISAT, Politecnico di Torino, Torino, Italy}
\affiliation{Italian Institute for Genomic Medicine, IRCCS Candiolo, Candiolo, Italy}

\author{Andrea Pagnani}
\affiliation{DISAT, Politecnico di Torino, Torino, Italy}
\affiliation{Italian Institute for Genomic Medicine, IRCCS Candiolo, Candiolo, Italy}
\affiliation{INFN, Sezione di Torino, Torino, Italy}

\author{Martin Weigt}
\affiliation{Sorbonne Universite, CNRS, Institut de Biologie Paris Seine, Computational and Quantitative Biology -- LCQB, Paris, France}

\begin{abstract}
This contribution focuses on the fascinating RNA
molecule, its  sequence-dependent folding driven by base-pairing interactions, the interplay between these interactions and natural evolution, 
and its multiple regulatory roles. The four of us have dug into these
topics using the tools and the spirit of the statistical physics of disordered systems, and in particular the concept of a disordered (energy/fitness) landscape.
After an introduction to RNA molecules and the perspectives they
open not only in evolutionary and synthetic biology but also in
medicine, we will introduce the important notions of energy and fitness
landscapes for these molecules. In Section III we will review some models and algorithms for
RNA sequence-to-secondary-structure mapping. Section IV discusses how 
the secondary-structure energy landscape can be derived from unzipping data. Section V deals with the inference of RNA structure from evolutionary sequence data
sampled in different organisms. This will shift the focus from the
`sequence-to-structure' mapping described in Section III to a `sequence-to-function' 
landscape that can be  inferred from laboratory evolutionary data on DNA
aptamers. Finally, in Section VI, we shall discuss the rich  theoretical picture linking networks of interacting RNA
molecules to the organization of robust, systemic regulatory programs. Along this path, we will therefore  explore phenomena across multiple scales in space, number of molecules and time, showing how the biological complexity of the RNA world can be captured by the unifying concepts of statistical physics.
\end{abstract}

\maketitle

\section{The diverse roles of RNA molecules}
\label{sec:rna_roles}
 Classical genetics in the middle of the 20th century consisted of a mere
 description of the passage of traits through generations, the
 physical underlying nature of ``Mendelian Genes'' being still unknown
 at the time. In 1944 O.~Avery proved that the ``transforming
 principle'' in aqueous cell-free extracts of pneumococci was
 desoxy-ribonucleic acid (DNA) \cite{avery43}. From this seminal
 discovery, the central role of the
 different nucleic acids became increasingly clear, along with their compartmentalization in the
 cell. While the eukaryotic nucleus hosted DNA molecules, proteins were
 synthesized in the cytoplasm and RNA molecules proved to be pivotal
 in this process \cite{brachet1956}. James Watson sketched the ``central
 dogma'' as early as 1952 (as anecdotically reported in  \cite{kornberg2002} and more precisely in \cite{crick1958}) proposing the now
 standard scenario of a protein-coding RNA (messenger RNA, mRNA) that is passed from the DNA to the
 protein synthetic machinery in the cytoplasm.

A second class of functional RNA was proposed in 1958 by Francis Crick in
his ``adaptor'' hypothesis \cite{crick1958}, positing the existence of a molecule
that translates the triplets of the genetic code into corresponding amino
acids. Interestingly, Crick suggested that not only such an adaptor
would be an RNA, but also that RNA would be a better evolutionary fit
over proteins as the material for his adaptors, because specific base-paring made it the ideal RNA-recognition molecule. His
intuition finally found an experimental proof in the discovery of
transfer RNAs (tRNA) by Mahlon Hoagland and co-workers
\cite{hoagland1958}.

RNA, originally considered in the ``central dogma'' as a mere
information carrying molecule, in the following decades has been found to have a
multitude of distinct functions. For instance, in protein synthesis, besides
the already discovered mRNAs and tRNAs, the structural and enzymatic role
of ribosomal RNA (rRNA) was identified. In the last two decades, many other forms of RNA have been discovered, including micro-RNAs (miRNA),
small interfering RNAs (siRNA), small nuclear RNAs (snRNA), small
nucleolar RNAs (sncsRNA), circular RNAs (circRNA), and long non-coding
RNAs (lncRNA). Notably, mRNA remains the only protein-coding RNA species discovered to date. Actually, the overwhelming majority of cellular RNA (by weight) in mammals is formed by non-coding molecules. This plethora of different non-coding RNAs performs
 diverse and sophisticated regulatory functions, some of them still poorly or not completely
understood: from helping to turn genes on and off to slicing  other RNA/DNA or transporting and ligating amino-acids. 

This versatility is nowadays in the spotlight of at least three different fields: synthetic biology, evolutionary biology and medicine.  In the field of synthetic biology, RNA-DNA based computing systems have been recently developed using DNA/RNA as a material capable of self-construction and communication and demonstrating its ability to perform adaptable logic gates \cite{kari1998dna,yurke2000dna}. 

 In evolutionary biology, its versatility makes RNA an
ideal candidate for playing the role of pre-Darwinian self-replicating
molecules, which predate the proper Darwinian DNA- and protein-based
evolution as we know it today. 
The so called {\em RNA world} hypothesis
\cite{gesteland1993}, posits a hypothetical stage of life on Earth, in
which self-replicating RNA molecules proliferate.
Virtually all biologists nowadays agree that bacterial cells cannot
form from nonliving chemicals in one step. If life arose from
nonliving chemicals, there must have been intermediate forms -- {\em
  precellular life} -- populating our pre-Darwinian Earth around 4
billion years ago. Since RNA may act both as a gene
(i.e. as a template genetic information storage) and as an enzyme, it bypasses the ``chicken-and-egg'' problem: genes require enzymes; and
enzymes require genes. As a last asset of the RNA world hypothesis, it is worth recalling that RNA can be (retro-)transcribed into DNA, in
reverse of the normal process of transcription.

Finally the diverse RNA roles have inspired ideas on how to use RNA in
medicine to develop RNA-based therapies, both using them as vaccines
against cancers and pathogens in form of mRNA templates for the synthesis of the
prescribed proteins, and as drugs to target and regulate nucleic acids
(RNA and DNA) and proteins \cite{deweerdt2019rna}.  In particular
several drugs have been developed based on oligonucleotides, small
stretches of single-strand DNA, made up of about 10-30 nucleotides,
which are used to prevent RNA from being translated into proteins by
several mechanisms including blocking the start of translation,
altering RNA splicing or tagging mRNA degradation.  Small RNA
or DNA molecules that target proteins, called aptamers, have been also
designed to bind a specific site of a specific protein to modulate its
function \cite{Damase2021}.

\section{From Fitness to Energy Landscapes}

The notion of landscape was introduced in biology in the late thirties by Sewall
Wright \cite{wright1931}, in order to describe evolution as an adaptive
walk in a fitness landscape.  Nowadays landscapes appear in so
different fields as physics of spin glasses, computer science of
combinatorial complexity, evolution, neural networks, gene regulatory
networks, maturation of immune response, and biophysics of
macro-molecules.  A geographical landscape can be described by a
height function defined on the two-dimensional Cartesian
plane. Here, more generally the notion of landscape relates to a
function $F({\bf x})$, which assigns a real number to all points ${\bf x} = (x_1,
x_2, \dots, x_n)$ of a $n$-dimensional space, also called 
configuration space. In geographical landscapes,
the configuration space is continuous, and the most natural metric is often
the Euclidean one. In the case of RNA and other sequence-based biological molecules, we frequently consider fitness landscapes defined on a discrete support.  We can more formally define a
landscape as a triple $(X,d,F)$, where the configuration space $X$ is a (possibly finite) set, $d : X
\times X \rightarrow {\mathbb R}^+$ is a metric, and $F : X
\rightarrow {\mathbb R}$ the landscape function. In the following we will assume that $F$ is a bounded-from-below energy/cost function, or bounded-from-above fitness function. Here the field-specific
definitions disagree: biologists follow Wright's original definition of maximum fitness, while physicists prefer to think in terms of
(free)-energy minimization. A simple minus sign relates the two.

It is instructive to think about biological landscapes in the context of
optimization theory, i.e.~in terms of a minimization problem of a
given cost function (sometimes internal energy or free
energy, sometimes negative fitness). Ever since Darwin, evolution was considered as the result of
the competition of mutation and selection: mutation acts on the
genotype (ultimately encoded in the DNA of the organism), while
selection on the phenotype. Their competition rules the complex
dynamics of evolution. It is evident that this point of view can hardly be
predictive without a suitable notion of energy or fitness function -- a daunting
task, which needs to take into account the distinct scales at which mutation and selection act, from the ``microscopic'' genetic scale up to all the complexity
involved for instance in organismal selection.

However, when we concentrate on bio-molecules, the situation becomes somewhat simpler. In particular
RNA makes a landscape-based modeling of sequence and structure
particularly promising: RNA is indeed a genetically active molecule
whose genotype (sequence) directly undergoes and inherits
mutations. Proteins, for instance, do not possess this
feature. Moreover, RNA secondary structure, thanks to its simplicity,
facilitates the task of defining a reliable cost function useful for
structural prediction from sequence - i.e.~of a biophysically motivated landscape function.

\section{RNA folding models}
RNA can be described via a hierarchy of
structures. The primary structure is the linear sequence composed by the four-letter RNA alphabet formed by the bases (or nucleotides) Adenine (A), Uracil (U), Cytosine
(C), and Guanine (G). As compared to DNA, the Thymine (T) in DNA is substituted by Uracil (U) in RNA. However, much like DNA, RNA can form stable
double helices of complementary strands. Since RNA usually occurs
single stranded, formation of double helical regions is accomplished
by the molecule folding back onto itself to form Watson-Crick (WC)
base pairs (G$\equiv$C and A=U), or the slightly less stable G--U wobble
pair (note that here the $\equiv,=,-$ symbols are  used just to give a rank of the interaction energies).
One of the most appealing features of RNA secondary
structure is that the obtained graph -- formed by the primary structure and the base pairs -- is planar as shown in
Fig.~\ref{fig:rnastruct}. Empirically, it turns out that this
planarity condition is often violated in particular in long structural
RNAs: so-called pseudo-knots (i.e. structural contacts
violating the planarity condition), are known to occur, although most
of the RNA secondary structure elements are planar. As is explained in
the following, the planarity condition allows for polynomial-time (in
sequence length) algorithms to compute the partition
function of a given RNA sequence. When, in particular as statistical
physicists, we try to describe a complex object such as an RNA
molecule, some kind of coarse-graining is necessary. As usual,
there are different languages for the different levels of
phenomenological descriptions. The secondary-structure one is
particularly interesting for the following reasons: (i)~conventional base paring and base pair stacking (viz. the stabilizing effect of sequences of ``consecutive'' base pairs $\{(p-k,q+k),(p-k+1,q+k-1),...,(p,q)\}$) provide the major part
of the free energy of folding; (ii) secondary structure has been used
successfully by biologists in the interpretation of RNA function and
activity; (iii) secondary structure is conserved from an
evolutionary standpoint. Physicists, computer scientists
and bioinformaticians have always been interested in RNA secondary
structures due to their discrete nature. RNA secondary structures are
easy to compare and to visualize. Lastly, and perhaps most importantly
here, the planarity condition allows for the implementation of
polynomial-time algorithms for the computation of
the native structure
\cite{nussinov1980,zuker1984,fontana1993,higgs1996,pagnani2000}. These algorithms use dynamic programming, i.e.~a recursive strategy starting from small solvable sub-problems and iterating towards larger and larger sub-problems, a technique resembling transfer matrices in statistical physics, or message passing / belief propagation in computer science.

\subsection{A mathematical model for the secondary structure description of RNAs}
The secondary structure of RNA is defined via a set of base pairs occurring in
its three-dimensional structure. Let us define a sequence as ${\cal R}
:= \{r_1, r_2, \dots, r_n\}$, with $r_i \in
\{A,C,G,U\}$ being the $i$-th base. A secondary structure on ${\cal R}$ is a set ${\cal S}$
of base pairs $(i,j)$, with the convention $1 \leq i \leq j \leq n$, fulfilling the following constraints:
\begin{enumerate}
    \item $r_i$ and $r_j$ form an allowed Watson-Crick or wobble pair;
    \item $j -i \geq 4$: this restriction permits RNA to loop back onto itself and to form antiparallel double-stranded stem helices;
    \item two distinct base pairs $(i, j), (i', j') \in {\cal S}$ are not nested: when assuming $i < i'$ without loss of generality, they fulfill one of the following conditions
    \begin{enumerate}
        \item $i < j < i' < j'$ : the pair $(i,j)$ precedes $(i',j')$, or
        \item $i < i' < j' < j$ : the pair $(i,j)$ includes $(i',j')$.
    \end{enumerate}
\end{enumerate}

\begin{figure}
\begin{center}
\includegraphics[width=0.8\columnwidth]{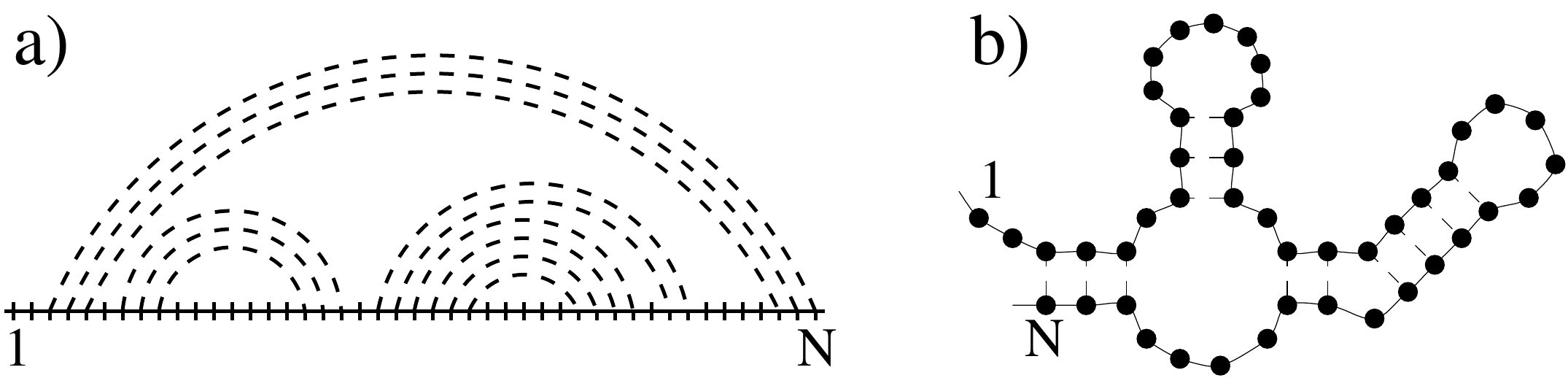}
\caption{Isomorphic representations of secondary structures in RNA: a) rainbow representation; b) cactus representation.}
\label{fig:rnastruct}
\end{center}
\end{figure}

The last condition defines a planar graph (i.e. the graph structure
can be drawn on a plane without link crossings) as shown in
Fig.~\ref{fig:rnastruct}. Although in nature there are examples of RNA structures which are knotted, we relate the existence of
knotted structures to the formation of tertiary structure (pseudoknots).

\subsection{Statistical mechanics of the secondary structure of random RNA sequences}

Knowing the fold of biomolecules like RNA and proteins
is crucial for understanding their biological functionality. Despite decades of active research to solve the so-called sequence-structure relationship, the computational prediction of biomolecular structure still remains a challenging problem. While there has been a major breakthrough in 2021 in the case of proteins \cite{jumper2021highly}, the problem remains widely open for RNA for what concerns in particular the prediction of tertiary (i.e.~the overall three dimensional arrangement of an RNA molecule)  and quaternary structures (i.e. the structure of an RNA molecule in complex with other biomolecules).

Real RNA sequences are not uniformly random; they show a remarkable sequence variability with subtle correlations that become visible
when homologous RNA sequences (i.e.~sequences deriving from a common ancestral sequence) are compared using multi-species multiple-sequence
alignments (MSA) \cite{nawrocki2013}.  The interest in studying the limiting
and somehow not biologically motivated case of really random sequences
arises from the need of answering the following question: is the
folding transition, that forces real biomolecules into their
functional shapes, characteristic of sequences selected by the
evolution? Or is it also present in random sequences? This issue is
particularly relevant in a prebiotic/pre-Darwinian era when, according
to the RNA-world hypothesis, sequences~\cite{gesteland1993} started
populating the environment, and, arguably, the self-replicating
selective pressure was acting on a pool of initially random RNA sequences.

\begin{figure}
\begin{center}
\includegraphics[width=0.9\columnwidth]{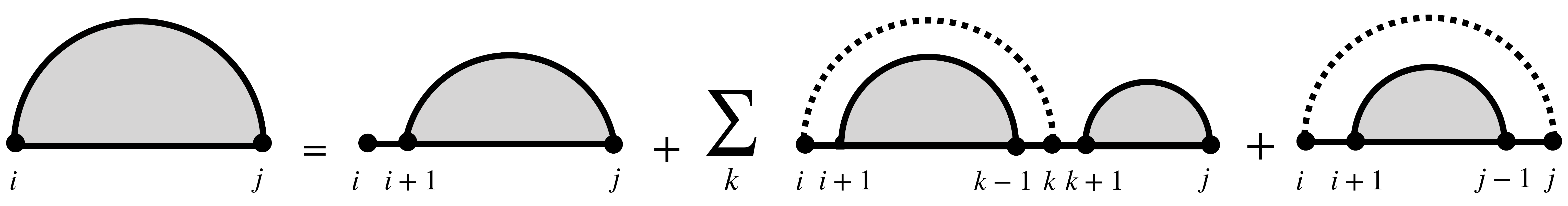}
\caption{Graphical representation of Eq.~(\ref{eq:zetarna}). We display as filled grey bubbles the partition function between
the two extremes of the bubble, and a term proportional to
$e^{-\beta e(r_i,r_j)}$ as a dotted semi-circular link. }
\label{fig_part_fun_rna}
\end{center}
\end{figure}

In our model, we introduce a drastic approximation. The energy of a structure is
simply defined as 
\begin{equation}
\label{ene_rna}
H[{\mathcal{S}}] = \sum_{(i,j)\in{\mathcal{S}}} e(r_{i},r_{j}) \ .
\end{equation}
Simple but reasonable values for $e(r_{i},r_{j})$ are $-3,
-2, -1 $ kcal/mole for G$\equiv$C, A=U, G--U base pairs respectively.  
More precise values of these parameters, taking into account the specific stacking energies of two consecutive base pairs, which depend on the  base pair and the following one, loop-conformational entropies, salt and temperature dependence~\cite{zuker1984}, have to be considered  to effectively describe the whole complexity of the energy landscape, e.g.~when fitting single molecule experiments, see Sections~\ref{sec::unzip} and~\ref{sec::disp}.
One of the advantages of this model is that
we clearly separate the role of disorder (encoded in the sequence
${\mathcal{R}}$) from that of frustration (induced by the
planarity condition on the structure ${\mathcal{S}}$). Thanks to the
planarity condition, it is possible to use a simple dynamic-programming algorithm to compute the partition function $Z_{i,j}$, which corresponds to the sum over all possible pairings between sites $i<j$:
\begin{equation}
Z_{i,j} = Z_{i+1,j}+ \sum_{k=i+1}^j Z_{i+1,k-1} e^{-\beta e(r_i,r_k)}
Z_{k+1,j} \quad ,
\label{eq:zetarna}
\end{equation}
as illustrated in Fig.~\ref{fig_part_fun_rna},
with $Z_{i,i}=Z_{i,i-1}=1$ for all $i\in 1,\dots,L$.

This recursion relation is particularly efficient since the time needed for
the computation of $Z_{1,L}$, i.e.~of the total partition function, scales as ${\cal O}(L^3)$. From the recursion,
we can compute explicitly many interesting quantities, such as the
probability that site $i$ and $j$ are linked~\cite{higgs1996}, integer
moments of the internal energy $U = \langle H^k[\mathcal{S}] \rangle$, and
the entropy.

\subsection{A glassy transition for disordered RNA models}

The model introduced in the previous subsection is a one-dimensional
model with long-range interactions. Numerically, it is indeed shown
that the probability of base pairing decays with the base pair distance with
an exponent $3/2$ at high-temperatures and somehow slower,
with an exponent near the value 4/3, at low temperatures~\cite{baez2019}.

A numerical study of the specific heat~\cite{pagnani2000} shows
clear signs of a freezing phase transition to a low-temperature glassy regime. From the sequence-to-sequence critical temperature fluctuations, a specific heat
critical exponent $\alpha \sim -1.9(1)$ is inferred; and the second
derivative of the specific heat with respect to the temperature should thus
display a very mild divergence or a finite jump. Near criticality,
the entropy of the model has a crossing point \cite{pagnani2000}, which
signals a rapid shrinking of the available phase space. Moreover, the
model has a finite zero-temperature entropy. A simple generalization
of the dynamic-programming equation, allows for an efficient exact
enumeration of all the ground-state structures (GSS) for any given
sequence. Since the model turns out to be highly degenerate in the
low-temperature phase, the natural question is how these GSS are
organized. It is quite obvious that a very different physical
behavior may appear in a model whose GSS are all very similar (like an
ordered or “ferromagnetic” behavior) compared to a model whose GSS
are dispersed over the whole configuration space. A more quantitative
analysis can be achieved introducing the notion of distance between
structures and a classification based on these distances. In order to
quantify the relative distance between two structures, the overlap of
secondary structures is defined as
\begin{equation}
  q_{{\cal S},{\cal S}'} = \frac{1}{L} \sum_{i,j} l^{({\cal S})}_{i,j} l^{({\cal S}')}_{i,j}\ ,
\end{equation} 
where the Boolean variable $l^{({\cal S})}_{i,j} $ is 1 if $i$ is
connected to $j$ in the structure ${\cal S}$, and zero otherwise. By
definition, the overlap takes values in the interval $[0, 1]$. For any
given disorder realization (i.e. sequence) ${\cal R}$, we define
the zero-temperature probability distribution function (pdf) of the
overlaps:
\begin{equation}
  P_{\cal R}(q) = \sum_{{\cal S},{\cal S'} \in \Gamma_{\cal R}} \delta(q-q_{{\cal S},{\cal S}'})\ ,
\end{equation}
with being $\Gamma_{\cal R}$ the GSS set. This definition can be easily
generalized to any temperature, by summing over all the structures
and weighting each term with its Gibbs-Boltzmann factor.  The usual
classification of disordered systems is based upon the average pdf of
the overlaps, the so-called $P(q) := \overline{P_{\cal R}(q)}$, where the
overline indicates a disorder average taken over RNA sequences. In the high-temperature phase, the
system is characterized by a simple paramagnetic $P(q)$
whose limiting behaviour for large $L$ is a delta function (the width of the
distribution goes to zero as $1/\sqrt{L}$). However, in the
low-temperature phase, the great majority of the sequences shows a
very broad $P_{\cal R}(q)$, signaling a strong heterogeneity in the
GSS. Moreover, the shape of the pdf becomes strongly sequence-dependent (viz. non self-averaging). Nevertheless, some common
features can be easily recognized: while single-peaked $P_{\cal R}(q)$ functions are mostly
associated with low-entropy sequences, highly structured $P_{\cal R}(q)$ functions do not seem to be correlated to the sequence's entropy, and they give rise to the broadness of the average $P(q)$.

The structural heterogeneity signalled by the broadness of the $P(q)$ for random sequences -- those that most likely dominated the early stage of prebiotic life -- has some interesting, albeit speculative, consequences for the RNA-world hypothesis. In principle two competing scenarios could have been possible:
\begin{itemize}
    \item A {\em ferromagnetic} scenario where all thermodynamically relevant structures look pretty much the same independently from the realization of the sequence.
    \item A (possibly effective) {\em replica-symmetry breaking} scenario, where structures are strongly dependent on the sequence realization.
\end{itemize}
From the results summarized above, the second scenario seems to be more likely. Of course, we have no hints on whether the structural richness displayed by this simple random RNA model is enough to maintain a prebiotic RNA world, the question being currently completely out of any experimental control. Still, the RSB scenario is a stimulating starting point.
 
\subsection{Secondary-structure prediction} 
 
The zero-temperature limit of Eq.~(\ref{eq:zetarna}) relates closely to one of the most prominent classical algorithms in RNA bioinformatics, the Nussinov algorithm\cite{nussinov1980} predicting the secondary structure for a given RNA sequence. The algorithm aims at finding the planary secondary structure which maximizes the number of realized Watson-Crick or wobble base pairs, i.e.~it corresponds to a simplified energy $e(r_i,r_j)=-1$ if $r_i$ and $r_j$ are complementary in the before-mentioned sense, and to $e(r_i,r_j)=0$ else.

The algorithm uses dynamic programming, and calculates iteratively the score $S(i,j)$ counting the maximum number of base pairs in the sub-sequence ${r_i,...,r_j}$. As an initialization, it starts from all sub-sequences of length zero and one, which naturally cannot have base pairs,
\begin{equation}
    \forall i: S(i,i-1) = S(i,i) = 0\ .
\end{equation}
In case of a constraint on the minimal loop length due to RNA stiffness, the initialization can be generalized easily.

As a next step, the zero-temperature limit of Eq.~(\ref{eq:zetarna}) is iterated. The sum over Boltzmann weights becomes a maximization:
\begin{equation}
    S(i,j) = \max 
    \left\{
    \begin{array}{ll}
    S(i+1,j)  &\  {\rm if }\ i\ {\rm unpaired}\\
    S(i+1,j-1) - e(r_i,r_j)  &\ {\rm if }\ i\ {\rm and } \ j\ {\rm paired}\\
    \underset{k\,|\,i<k<j}{\max}  S(i,k) + S(k+1,j) &\ {\rm if}\ 
    i\ {\rm paired\ with\ internal}\ k .
    \end{array}
    \right.
\end{equation}
The final score $S(1,L)$ gives the maximum number of possible base pairs formed by the sequence $\{r_1,...,r_L\}$ within a secondary structure. To get this secondary structure, we have to trace back the algorithm to extract the path, which realized the final maximum score.

While this algorithm is pretty efficient -- it predicts the secondary structure in time ${\cal O}(L^3)$ -- the accuracy of the prediction is very limited. This can be improved along a number of lines, in all cases maintaining the basic algorithmic structure of the Nussinov algorithm: first, instead of counting base pairs, we can minimize their free energy, also taking into account the different binding energies of the different base pairs, the stacking of base pairs in stems, and the conformational entropy of loops; this is implemented in particular in the Vienna package\cite{lorenz2011viennarna}. Second, we can determine the consensus secondary structure of a MSA of homologous RNAs. This will be discussed in more detail below.

\section{RNA unzipping: from experiments to theory }

\begin{figure}
\begin{center}
\includegraphics[width=1\columnwidth]{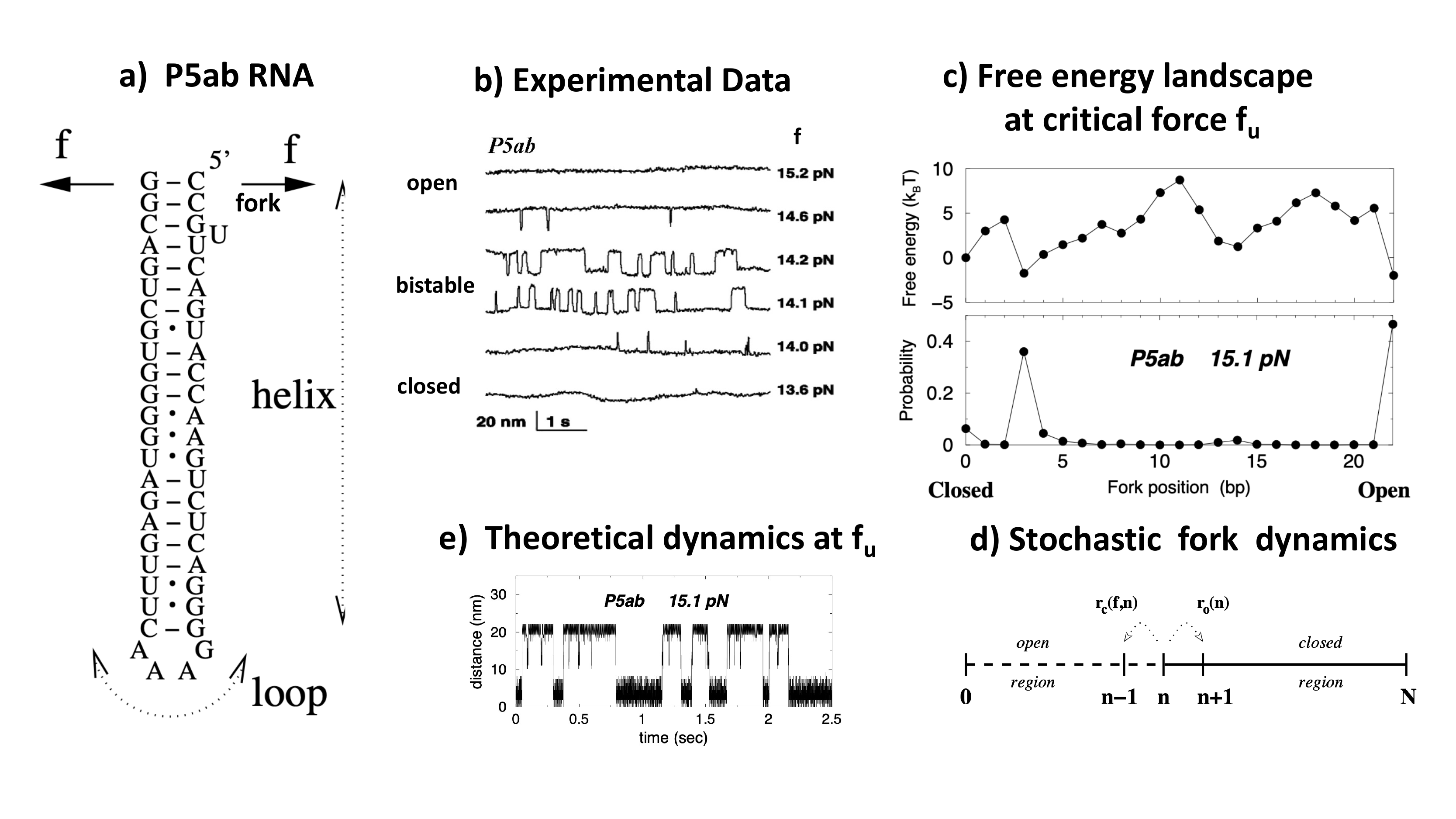}
\caption{a) Sketch of the unzipping by a constant force of the P5ab  hairpin.
b) Experimental  unzipping signal given by the distance between the two extremities of the RNA as a function of the time for different forces. For $f<f_u$ the molecule stays closed, at $f=f_u$ a bistable signal is observed, at $f>f_u$ the molecule stays open. c) The sequence dependent unzipping free energy landscape at force $f_u$ shows three minima in 0, 3 and 21 opened base pairs, on which the  configurational probability is peaked. d+e) A stochastic dynamics for the opening force in the unzipping landscape reproduces the unzipping signal. The figure is re-adapted from Ref.~\cite{cocco2003slow}. }
\label{fig:P5ab}
\end{center}
\end{figure}
In all regulatory processes mentioned in Sec.~\ref{sec:rna_roles}, the 
RNA complementary pairing plays a key role. For example, in therapeutic aptamers  designed as anticoagulants  that bind the  thrombin protein thus avoiding its auto-aggregation, a simple hairpin structure is present. The sequence in the loop structure is designed to bind thrombin \cite{zhou2019dna}.
In other cases,  such as the  specific  recognition and binding in the  CRISPR-Cas9 gene editing system \cite{jinek2012programmable}, RNA-DNA pairing is at the basis of the direct interaction with the DNA to be cut. Finally, both direct  pairing and loop binding mechanisms are present in
 the regulatory RNAs called riboswitches, which upon binding to metabolites,  control gene expression \cite{vitreschak2004riboswitches}. Riboswitches are aptamers characterized by structures made by few helix-loop units (see Fig.~\ref{fig:rnastruct} and Sec.~\ref{sec:dca}). The interaction with the metabolite happens on an unpaired nucleotide and brings to an alternative secondary structure, in which the riboswitch directly pairs to the expression platform of the nearby gene and turns it on and off.

In the natural regulatory RNA described in the  examples above, the pairing has to be strong enough to make the regulatory process happen, but it also has to  be transitory, to allow to stop the regulatory interaction and to enable a different operation to be carried out on the gene. We are just starting to understand the importance and complexity of such dynamical pairing/unpairing processes  and how to engineer DNA/RNA oligonucleotides able to have desired secondary structure and dynamical properties, such as multi-stable desired  structures. Achieving such control could  open  the field to the bio-engineering of RNA/DNA regulation processes and DNA bio-computing \cite{yurke2000dna}.  

Since the 1990s, single molecule manipulations \cite{bockelmann1997molecular} have made possible the direct investigation of the DNA pairing/unpairing dynamics. Thanks to optical and magnetic traps, single RNA and DNA molecules can be stretched or unzipped, by separating the complementary strands and breaking the base pairing between them; interactions with other nucleic acid or proteins can also be investigated \cite{marko2003micromechanics,bustamante1994entropic,cluzel1996dna,allemand1998stretched,tinoco2006determination}. 
 A more detailed description of single-molecule experiments by optical traps is given in the following contribution by F.~Ritort. 
 
 In the following, we describe how the statistical physics of random walks in a disordered energy landscape -- with the disorder again introduced by the RNA sequence -- has contributed to the interpretation of single molecule experiments and the general understanding of pairing/unpairing processes in RNA \cite{tinoco2006determination,lubensky2000pulling,cocco2002theoretical}.
 We describe some works we have carried out to reproduce the unzipping dynamics of simple RNA hairpins \cite{cocco2003slow} and the displacement interactions between  oligo-nucleotides bound on a DNA strand and the DNA closing hairpin  \cite{Ding2022.04.30.490171}. The inverse  problem of inferring the RNA and DNA sequence for unzipping experiments \cite{baldazzi2006inference} will be also mentioned.   

 \subsection{Unzipping and RNA secondary structure}
 \label{sec::unzip}
 The free energy to unzip the first $n$ out of $N$ base pairs in a RNA hairpin at constant force $f$ (cf.~Fig.~\ref{fig:P5ab}a) reads
\begin{equation}
\label{ene_rna_0}
H[{\mathcal{R}},f,n] = -\sum_{a=1}^{n} \left\{ e(r_{i(a)},r_{i(a+1)},r_{j(a)},r_{j(a+1)}) - 2 n g_{s}(f,a) \right\} \ ,
\end{equation}
where the pairing free energy parameters $e_a = e(r_{i(a)},r_{j(a)},r_{j(a)},r_{j(a+1)})$ depend on the identity of the $a$th base pair $\{i(a),j(a)\}$, $1\leq a\leq N$, and on its stacking with the $(a+1)$st base pair (if adjacent), since both base pairing and stacking need to be broken by the applied force. They are determined by thermodynamic measurements. The parameters taken from, e.g., RNAfold \cite{lorenz2011viennarna} were obtained by thermodynamic calibration. The negative $2 g_{s}(f,a)$ is the stretching energy, gained when opening the $a$th base pair \cite{cocco2003slow}. Polymer physics models  have been used to describe such elastic response and the parameters have been calibrated by single molecule experiments \cite{cocco2002theoretical,marko2003micromechanics}.
An unzipping force of  about $15 pN$  compensates the loss in pairing energy (of the order of $ -2.5 k_B T$) with the gain in stretching free energy of the opened base pair. This value gives the typical critical unzipping forces at which the molecule starts to open. 

The unzipping free-energy landscape  of a homopolymer  is  flat at the critical unzipping force: every configuration of $n$ open base pairs is equally probable and the opening fork makes a diffusive motion between a closed and an open molecule.  However for a heteropolymer like RNA, a free energy landscape with barriers and minima determines a  characteristic opening-closing time for the molecule, much longer than the  diffusive time.
As is shown in Fig.~\ref{fig:P5ab}c for the simple P5ab RNA hairpin, and generally valid for all hairpins with stem-loop structures, large barriers are still present at the critical unzipping force. This is due to the fact that  when opening a single unpaired base (mismatch) or a region of them (loop) there is no cost in pairing energy,  the contribution of this region to the opening free energy is therefore negative.

The P5ab free-energy landscape \cite{cocco2003slow} is characterized by three minima, in which the opening fork has the largest probabilities to be located. The telegraphic signal experimentally observed \cite{liphardt2001reversible}, see Fig.~\ref{fig:P5ab}b, corresponds to the persistence in these states and jumps between them. This signal can be precisely described by a stochastic dynamics of the opening fork in the above free energy landscape with opening  $R_o (n)$ and closing $R_c (f,n) $ rates,
 \begin{equation}
\label{rate}
R_o (a)= R\; e^{e_a/k_B T} \ , \qquad
R_c (f,a)=R \;e^{2 g_s(f,a)/k_B T} \ ,
\end{equation} 
where $R$ is the microscopic fluctuation rate for a base pair;
it was estimated to be $R \approx 5\times 10^6 s^{-1}$ by computing
the inverse self-diffusion time for a few-nanometer diameter object\cite{doi1988theory}.

The rates in Eq.~(\ref{rate}) lead to a master equation for the probability
$\rho _a (t)$ for the opening fork to be at site $a$ at time~$t$,
\begin{equation}
\label{master1}
\begin{split}
\rho _a (t+\delta t) &= \rho _a(t)+ \delta t \left[ -\rho _a(t)  \left(
R_o(a)+R_c(f,a)\right) \right. \\ &+ \left. \rho _{a-1}(t)  R_o(a-1) 
+\rho_{a+1}(t)  R_c(f,a+1) \right]
\end{split}
\end{equation}
or, equivalently, in the time continuum limit,
\begin{equation}
\label{master}
\frac{d\, \rho _a(t)}{d\,t}=-\sum_{b=0}^{N}\; T_{a,b}\; \rho_b (t)\ .
\end{equation}
This $(N+1) \times (N+1)$  matrix $T_{a,b}$ is tridiagonal, corresponding to opening one base pair at the time, with nonzero 
entries $T_{a-1,a}=-R_c(f,a)$, $T_{a+1,a}=-R_o(a)$, and 
$T_{a,a}=R_c(f,a)+ R_o(a) $.
The characteristic switching time can be computed  through the smallest non-zero eigenvalue of the transition matrix, $t_{switch}= 1/(\lambda_1 R) $ 
with $\lambda _1= 2 \times 10^{-6}$ for P5ab, which is well separated from the others. By fitting the value of $R = 3.6 \times 10^6 \; s^{-1}$  one finds back the observed switching time $t=0.14 s$.
 
It is possible to reverse engineer the problem to design hairpin structures displaying bistable dynamics on a desired time scale.
For a uniform sequence this can be easily achieved  just by acting on the loop and hairpin lengths. Consider a constant pairing free energy $e_0$ for each base pair. The free energy barrier $G^*$ at criticality for a 
$N$-base-pair RNA stem (i.e.~double helix)
followed by a $L$-base
loop (closing free energy $g_{loop}(L)$ at zero force) (Fig.~\ref{fig:P5ab}) can
be  easily estimated. The critical force $f_u$
is given by the condition that the free energy of the open molecule
equals the free energy  of the closed molecule:
$G(0,f_u)=G(N,f_u)$, that is, 
$0=- N \, e_0 + (2 N+L)\,g_{ss} (f_u)+ g_{loop} (L)$.
The barrier height $G^* ( N,L)\equiv G(N-1,f_u)$ then reads
\begin{equation}
G^*(N,L)=  (N -1) (- e_0+2 \,g_{ss} (f^*) )=\frac{(N-1) 
\left( -e_0\; L -2 \;g_{loop}(L) \right)}
{L+2 N} \ .
\label{steam}
\end{equation}
As sketched in the introduction, the possibility of bio-engineering switching times is important in the field of DNA/RNA-computing to create dynamical programming with this elementary processes \cite{wang2020implementing,yurke2000dna}.
Another interesting application of DNA/RNA unzipping is in sequencing technologies,  by reconstructing the amino-acid sequence from the unzipping energy landscape and its heterogeneities due to the sequence \cite{bockelmann1997molecular,baldazzi2006inference,baldazzi2007inferring,barbieri2014reconstruction}.
Sequencing through unzipping could be advantageous with respect to existing sequencing technologies by avoiding short reads  which make difficult, e.g., the correct sequencing of repeated sequences.  
Fluctuations of the open single strand have so far made it difficult to practically use this idea to sequence long molecules. 
The problem of reconstructing the sequence from the opening signal is, however, a very interesting theoretical problem, corresponding to the inverse problem of the  Sinai model with a drift describing random walks in a disordered free energy landscape tilted by a force \cite{cocco2007reconstructing,cocco2009inference}. Applications of the inverse problem have been realized in nanopore sequencing, in which the sequence is unzipped by passing through a nanopore channel: from the detected electrical signal, the unzipped base pair is reconstructed \cite{MATHE20043205}. 
 
 \subsection{Oligonucleotides displacement models}
 \label{sec::disp}
 
 \begin{figure}
\begin{center}
\includegraphics[width=1\columnwidth]{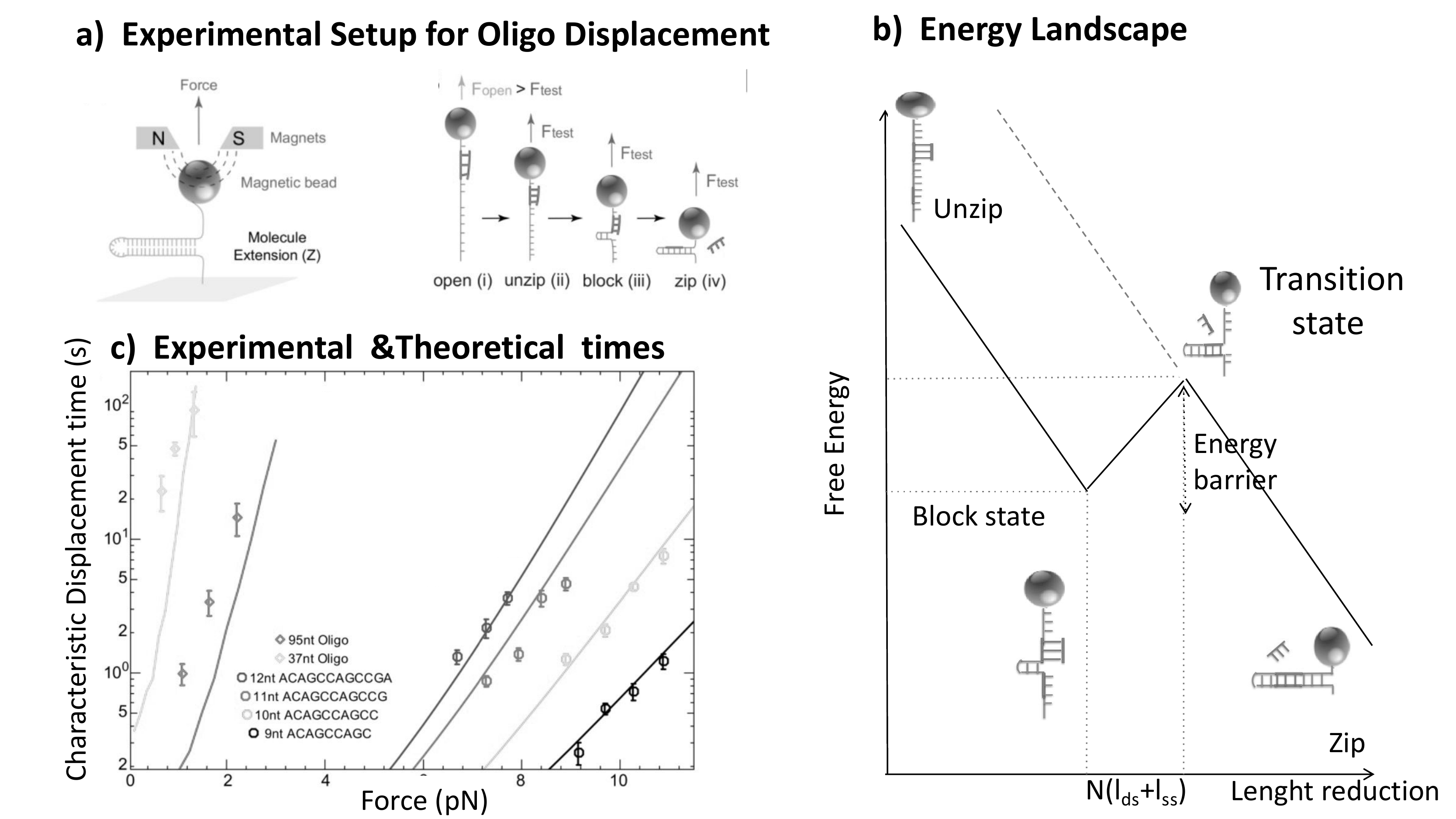}
\caption{a) Experimental Setup for the single-molecule oligonucleotide displacement process. The complementary DNA strand is first unzipped at large unzipping force ($f>15 pN$)  and an oligonucleotide is let to hybridize on its complementary segment. The force is then lowered to the test force $f_{test}<15 pN$, and the DNA strand starts to close at such force until the closing strand hits the oligonucleotide, which has to be displaced before continuing the closing process. This obstacle generates a barrier in the energy landscape (b). c) Measured and theoretical characteristic displacement times as a function of the test force for oligonucleotides of sizes between 9 and 95 nucleotides. The figure is re-adapted from Ref.~\cite{Ding2022.04.30.490171}.}
\label{fig_displacement}
\end{center}
\end{figure}

 In the following, we will illustrate how single molecule experiments and the RNA/DNA opening free-energy landscape can be used to predict for how long and where a probe made by a complementary segment of DNA can be bound to a DNA strand, before dissociating naturally or being displaced by another DNA. More precisely, in the experiment described below (Fig.~\ref{fig_displacement}), the closure of a double-stranded DNA can be blocked by the hybridization of the complementary probe, and the position and the blocking time can be monitored and theoretically predicted. This technique can be used in single-molecule sequencing~\cite{wang2021detection} to find a particular sequence on a DNA molecule. Moreover, monitoring the blocking time allows one to detect the presence of a mutation or a mismatch between the probe and the DNA sequence,  typically giving shorter blocking times.
 The  unzipping free-energy landscape Eq.~(\ref{ene_rna}), at room temperature and zero unzipping force, 
 quantifies the  cost  to spontaneously  dissociate a DNA or RNA molecule from a DNA substrate.
 For each base pair to break, the increase of free energy of about $2.5 k_B T$ multiplies the opening time by a factor 10. It is therefore impossible to observe the spontaneous dissociation  of a DNA/RNA probe of more than a dozen base pairs from its complementary DNA/RNA strand. This can be a serious limitation not only in vitro to monitor the blocking time, but mainly in vivo for the regulatory processes RNA has to carry out on DNA genomes, cf.~the introduction,  as well as for the bio-engineered imitations of such processes.
 One way out is to exploit the base-by-base displacement process: it has been shown by single molecule experiments that this is a very efficient and  powerful mechanism to dissociate molecules, not only for DNA/RNA oligonocleotides but also for proteins \cite{graham2011concentration,tang2020dna,Ding2022.04.30.490171}, which are otherwise very stably bound to the complementary DNA. The replacement process happens thanks to thermal fluctuation by replacing one component of the bound molecule at the time \cite{cocco2014stochastic} and can dissociate, in few minutes, DNA oligonucleotides  as long as 95 bases, as observed by the single molecule experiments \cite{Ding2022.04.30.490171} described in Fig.~\ref{fig_displacement}. Similarly to unzipping, a model describing the random motion of the opening fork on the displacement landscape (cf.~Fig.~\ref{fig_displacement}b) can be introduced to mathematically describe the replacement process at any unzipping force $f$ (even very small). The predicted displacement times as a function of the length of the complementary oligonucleotide and its sequence are in good agreement with experiments, as is shown in Fig.~\ref{fig_displacement}c. Differently from Eq.~(\ref{ene_rna}), the energy barrier for the progression of the displacement
 fork does not include any pairing energy, as the same pairing energy lost for the bound oligonucleotide is gained by  closing the DNA-hairpin.  
 The transition matrix becomes slightly more complex than the one for unzipping, Eq.~(\ref{master}), as it needs to take into account the  hairpin fork and the two  oligonucleotide ends. Moreover,  configurations  corresponding to both the closing DNA-hairpin and the oligonucleotide  pairing on the same base exist. Still it is possible to numerically diagonalize this matrix and to obtain very accurate estimates of the replacement times, cf.~Fig.~\ref{fig_displacement}. Interestingly the displacement times are not strongly affected  by changing the pairing parameters, as they only marginally depend on pairing energies. On the contrary, as soon as there is an asymmetry between the bound molecule and the invader, e.g.~for hybrid  RNA-DNA molecules and for mismatches, the displacement times become very sensitive to the sequence and mismatch position.

\section{RNA evolution, coevolution, and statistical sequence models}
\label{sec:dca}

\subsection{RNA families and (co)evolutionary constraints}
 
As all biological molecules, RNAs are subject to natural evolution. Nucleotide mutations introduce random changes into the RNA sequence. But not all changes lead to functional variants, and natural selection tends to suppress deleterious mutations. In the case of functional RNA two selective constraints are of particular importance: (i) some functionally important motifs contains combinations of positions in the sequence, which cannot be changed, they are highly conserved, (ii) functional RNA tend to conserve their secondary structure, realized via the complementarity of the concerned base pairs. If, e.g., the first nucleotide in the base pair C-G is mutated into an A, the pairing is disrupted, and the corresponding secondary structure weakened. However, if also the second position is changed into a U, we find an alternative base pair A-U. The two positions thus cannot evolve independently, they {\em coevolve} \cite{levitt1969detailed,fox19755s}.

\begin{figure}
\begin{center}
\includegraphics[width=0.9\columnwidth]{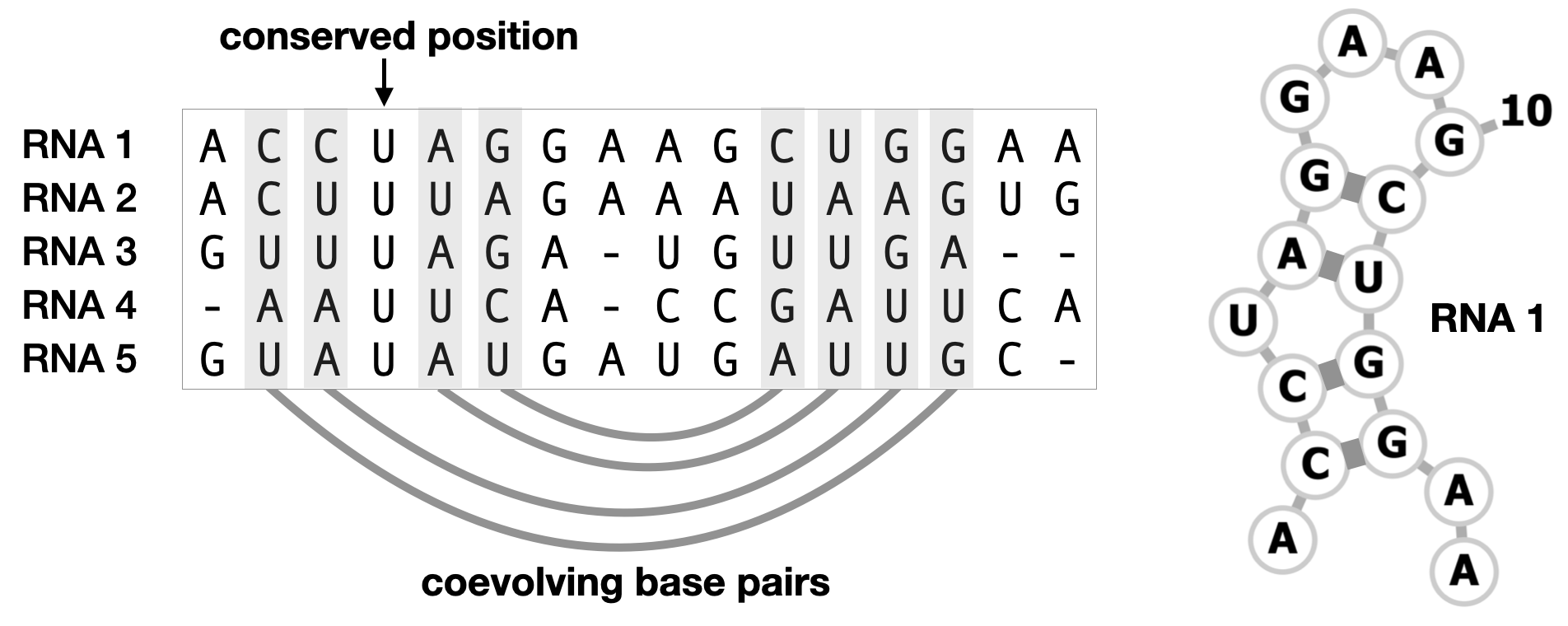}
\caption{Coevolution of base pairs in RNA secondary structure, as represented in a small MSA of RNA sequences. The evidenced pairs of columns are coevolving. While the individual nucleotides change from sequence to sequence, the pairs of same color are compatible with base pairing (Watson-Crick or wobble pairs). All sequences are compatible with the secondary structure depicted on the right (nucleotides labeled according to the first sequence, coevolving secondary-structure pairs connected by bold lines). }
\label{fig:rna_msa}
\end{center}
\end{figure}

Databases like Rfam \cite{kalvari2021rfam} collect families of homologous RNA sequences, i.e.~of sequences having common ancestry in evolution, having typically very similar secondary structure and biological function, but potentially distantly diverged sequences. These families are provided in the form of {\em multiple-sequence alignments} (MSA), i.e. in the form of rectangular matrices, where each row corresponds to an individual RNA sequence, and each column to a nucleotide position resulting from shared evolutionary ancestry, cf. the schematic representation in Fig.~\ref{fig:rna_msa}. The problem of aligning multiple RNA sequences is non-trivial: first, sequences may change their length via nucleotide insertions or deletions, therefore MSA do not only contain the four nucleotides A, C, G and U, but also the alignment {\em gap} ``--'' accounting for missing nucleotides. A gap may indicate a deletion in the sequence, or an insertion into another sequence. Second, in the case of base pairs, the individual nucleotides may be highly variable, but their complementarity needs to be conserved. In difference to standard MSA techniques developed for DNA or  protein sequences\cite{eddy2011accelerated}, algorithms for RNA sequence alignment need to take this coevolutionary information into account \cite{nawrocki2013}.

But how much does the conservation of the secondary structure restrict the viable RNA sequence space? Advanced MCMC sampling techniques are used\cite{jorg2008neutral} in combination with secondary structure prediction tools \cite{gruber2008vienna} to estimate the size of the so-called {\em neutral network} \cite{schuster1994sequences} in RNA sequence space, defined as the collection of all nucleotide sequences, which would fold into the same secondary structure. These studies found an impressive reduction in viable sequence numbers: to give an example, for RNA sequences of 200 nucleotides, the sequence space is reduced from a total of $4^{200}\simeq 2.6\times 10^{120}$ sequences to ``only'' about $7\times 10^{86}$ sequences being predicted to have the correct secondary structure. While this corresponds to a reduction of over 30 orders of magnitude, the resulting neutral network is still astronomically large (for comparison, there are about $10^{80}$ atoms in the entire universe).

So, it becomes clear that evolutionary constraints – conservation and coevolution  play a crucial role on RNA evolution and should leave important statistical traces in MSA of homologous sequences. However, even the largest families sample only a tiny fraction of the viable sequence space. The tRNA family has currently 1.4 million sequence entries in Rfam, being by far the largest RNA family. All other families have (in most cases much) less than 200,000 sequences. Statistical analysis or modeling of RNA sequences therefore works in the regime of extreme undersampling - despite what is called the ``data revolution'' or ``sequencing revolution'' in modern biology.

\subsection{Profile and covariance models}

Assume an MSA ${\cal M} = (r_i^\mu)_{i=1,...,L}^{\mu=1,...,M}$ of $M$ homologous RNA sequences of aligned length $L$ to be given. The aim of statistical sequence models is to represent the statistical properties of these RNA sequences via a statistical model $P(r_1,...,r_L)$ assigning a probability to each possible nucleotide sequence over the extended alphabet \{A,C,G,U,--\} containing the four possible nucleotides and the alignment gap. In analogy to statistical mechanics, we will represent these statistical models as Boltzmann distributions $P(r_1,...,r_L)\sim \exp\{-H(r_1,...,r_L)\}$ introducing the {\em statistical energy} $H(r_1,...,r_L)$. This function has to be learnt from the MSA ${\cal M}$.

Two remarks are in order: first, the model resembles the energy introduced in Eq.~(\ref{ene_rna}), but it has a totally different meaning. While Eq.~(\ref{ene_rna}) describes a statistical-physics model for the secondary structure at {\em given} RNA sequence, the random variable in $P(r_1,...,r_L)$ (or $H(r_1,...,r_L)$) is the sequence itself. Second, since we have to learn our statistical energy from data in the case of extreme undersampling, we need to assume some suitable parameterization of $H$. We need to keep in mind that too simple parameterizations will not be able to represent the statistics of the MSA, while too complex models may suffer from overfitting, i.e. from fitting the noise contained in the specific finite sample given by the MSA. In the bioinformatics of RNA sequences, two classes of statistical models are intensively used \cite{durbin1998biological}. 

The simplest non-trivial class of models are {\em profile models} \cite{durbin1998biological}. In their simplest form, their statistical energy takes the form
\begin{equation}
    \label{eq:profile}
    H(r_1,...,r_L) = -\sum_{i=1}^L h_i(r_i)
\end{equation}
of a sum of position-dependent local ``fields'' (or biases), i.e. the model $P(r_1,...,r_L)$ describes $L$ statistically independent positions. A simple maximum-likelihood approach allows to relate the fields to the statistics of single MSA columns. Let us consider $f_i(r)=\frac 1M \sum_\mu \delta_{r,r_i^\mu}$ the fraction of sequences in ${\cal M}$ having nucleotide (or gap) $r$ in position (or MSA column) $i$. The local fields are then directly given by the empirical log-frequencies,
\begin{equation}
    \label{eq:fields}
    h_i(r) = \ln f_i(r) \ .
\end{equation}
 Note that profile models can capture patterns of nucleotide conservation due to the position-specific fields: a nucleotide frequently observed in a position will have a more favorable local field than a rare nucleotide. However, due to their factorized form, profile models cannot capture coevolution. Nevertheless they belong to the most successful statistical models in sequence bioinformatics, in particular for DNA and protein sequences, and are at the basis of most approaches to sequence alignment, homology detection or phylogeny reconstruction \cite{durbin1998biological}.

We motivated coevolution by the need to conserve secondary structure in RNA evolution. In a base pair, connecting two columns $i$ and $j$ in our MSA, we will therefore find more Watson-Crick or wobble pairs A-U, U-A, G-C, C-G, G-U and U-G than any other nucleotide pair. This means that the joint nucleotide frequency $f_{i,j}(r,r')=\frac 1M \sum_\mu \delta_{r,r_i^\mu} \delta_{r',r_j^\mu}$ of nucleotide pairs $(r,r')$ in positions $i$ and $j$ deviates strongly from the factorized expression $f_i(r) f_j(r')$. This fact is taken into account in {\em covariance models} for RNA sequence ensembles at given secondary structure ${\cal S}$ \cite{eddy1994rna}. Their statistical energy reads
\begin{equation}
    \label{eq:cm}
    H(r_1,...,r_L|{\cal S}) = -\sum_{i=1}^L h_i(r_i) - \sum_{(ij)\in {\cal S}} J_{ij}(r_i, r_j)\ ,
\end{equation}
i.e.~covariance models introduce explicit pair couplings for all base pairs in ${\cal S}$. Also here the simple graphical structure of ${\cal S}$ allows us to analytically find the maximum-likelihood parameters for fields and couplings: sites are at most included in one base pair, and Eq.~(\ref{eq:cm}) describes a collection of individual sites and paired dimers, which are statistically independent between each other. While the fields are still given by Eq.~\eqref{eq:fields}, the couplings read
\begin{equation}
    \label{eq:J}
    J_{ij}(r,r') = \ln \frac {f_{i,j}(r,r')}{f_i(r) f_j(r')}
\end{equation}
for all $(ij)\in {\cal S}$: they are given as the log-ratio between the joint empirical frequency and the factorized expression. Covariance models, while still being easily treatable, greatly enhance the accuracy of RNA-sequence alignment and homology detection \cite{nawrocki2013}.

\subsection{Coevolution-based secondary structure prediction}

Using a covariance model given by a secondary RNA structure ${\cal S}$ and the maximum-likelihood parameters in Eqs.~(\ref{eq:fields}) and (\ref{eq:J}), and assuming that the sequences in the MSA ${\cal M}$ are independently drawn from each other (an assumption typically violated by sequence data characterized by phylogenetic correlations), we can calculate the log-probability of the MSA:
\begin{eqnarray}
    \label{eq:logprob}
    \ln P({\cal M} | {\cal S}) &=& \frac 1M 
    \sum_{\mu=1}^M \ln P( r_1^\mu,...,r_L^\mu | {\cal S})
    \nonumber\\
    &=& -\sum_{i=1}^L s_i + \sum_{(ij)\in {\cal S}} MI_{ij}\ ,
\end{eqnarray}
where we have introduced the empirical single-site entropies
\begin{equation}
    \label{eq:s}
    s_i = -\sum_r f_i(r) \ln f_i(r)
\end{equation}
characterizing the variation (or conservation) of individual positions, and the mutual information
\begin{equation}
    \label{eq:MI}
    MI_{ij} = \sum_{r,r'} f_{ij}(r,r') \ln 
    \frac {f_{i,j}(r,r')}{f_i(r) f_j(r')}
\end{equation}
measuring the covariation (or coevolution) of pairs of positions. Following standards in inference, we can use this expression as the log-likelihood of the secondary structure ${\cal S}$ given the data ${\cal M}$, which up to ${\cal S}$-independent terms reads
\begin{equation}
    \label{eq:ll}
    {\cal L} ({\cal S} | {\cal M}) = \sum_{(ij)\in {\cal S}} MI_{ij} + {\rm const} \ ,
\end{equation}
and find the best possible secondary structure by maximum likelihood. Note that again this has a very striking similarity to the single-sequence energy in Eq.~(\ref{ene_rna}), but with the physical interactions between nucleotides replaced by the mutual information between nucleotide positions (or MSA columns). It is now pretty straightforward to generalize the Nussinov algorithm for single-sequence secondary-structure prediction to the prediction of a {\em consensus secondary structure} of the MSA, by replacing the number of base pairs with their cumulative mutual information. This maximizes the likelihood of ${\cal S}$ for a given MSA over all non-nested planar secondary structures.

In RNA bioinformatics, it is well known that single-sequence secondary-structure prediction is imprecise. Therefore many current secondary-structure prediction tools use MSA and pairwise mutual information in their core algorithms \cite{lorenz2011viennarna}.

\subsection{Direct-Coupling Analysis for RNA families}

Proceeding in this way, we need to first calculate the mutual information $MI_{ij}$ for all pairs $(ij)$ of positions, i.e. for all $1\leq i<j\leq L$. Out of these $\binom{L}{2}$ pairs, less than $L/2$ can be part of the base pairs of the secondary structure. Analyses of MI-values in biological MSA show that not only these base pairs coevolve, but there are further pairs showing considerable correlation in nucleotide usage. Some of them correspond to neighboring positions along the sequence (primary structure), some to the before mentioned pseudo-knots or other three-dimensional contacts between nucleotides, others lack a simple interpretation.

It seems therefore wise to take into account in the statistical modeling of an RNA family also the coevolution between nucleotide pairs not paired in the secondary structure. In the absence of a clear selection criterion for important pairs, we follow the idea of the {\em Direct-Coupling Analysis} (DCA) \cite{morcos2011direct,cocco2018inverse} also explained more theoretically and for proteins in the chapter by Aurell et al. We construct a {\em Potts model} (aka Markov Random Field or Boltzmann machine) given by the statistical energy
\begin{equation}
    \label{eq:rna_dca}
    H(r_1,...,r_L) = -\sum_{1\leq i \leq L} h_i(r_i) - \sum_{1\leq i<j<L} J_{ij}(r_i, r_j)\ ,
\end{equation}
which is characterized by a fully connected network of -- a priori fully disordered -- pairwise couplings. In practice, these couplings are disordered but (up to some gauge or reparameterization invariance) determined by the conditions
\begin{eqnarray}
    \label{eq:marginals}
    \sum_{r_1,...,r_L} P(r_1,...,r_L) \delta_{r,r_i} & = & f_i(r) \ , \nonumber\\
    \sum_{r_1,...,r_L} P(r_1,...,r_L) \delta_{r,r_i} \delta_{r',r_j} & = & f_{ij}(r,r')\ ,
\end{eqnarray}
i.e. all one- and two-site marginals of $P$ have to coincide with the corresponding empirical nucleotide frequencies. Obviously, the determination of the model parameters – fields and couplings – now becomes a computationally hard task, since the exact calculation of the marginals of a given models requires the sum over $5^L$ possible aligned nucleotide sequences. 

The computationally simplest but rather time-demanding way is to estimate the left-hand site of Eqs.~(\ref{eq:marginals}) using MCMC simulations, and to iteratively update the model parameters to reach equality with the empirical frequencies. While this procedure, known as Boltzmann-machine learning \cite{ackley1985learning}, leads to very precise models, in many tasks much faster but less accurate approximations can be used. The chapter by Aurell et al. describes in particular the widely used mean-field \cite{morcos2011direct} and pseudo-likelihood maximization \cite{balakrishnan2011learning,ekeberg2013improved} approaches, in particular the first one being motivated by prior work on high-temperature expansions in spin-glass systems \cite{plefka1982convergence,georges1991expand}.

Once model parameters are inferred, we can use them to extract biologically important information \cite{de2015direct,weinreb20163d}:
\begin{enumerate}
    \item {\em Secondary-structure prediction:} Instead of using mutual information for secondary-structure prediction as discussed in the last subsection, we can use the statistical couplings $J_{ij}$. This was found to slightly improve the predicted secondary structure.
    \item {\em Tertiary-structure prediction:} More importantly, the couplings may contain information going beyond secondary structure, they may describe also tertiary-structure contacts. While it was found that this signal is rather weak as compared to the strong base-pair coevolution, it is still able to help for tertiary structure prediction.
    \item {\em Evolution-guided sequence design:} Accurately inferred DCA models (using Boltzmann machine learning) were found to be generative \cite{cuturello2020assessing}. This means that sequences sampled from the model using MCMC are statistically hardly distinguishable from natural RNA sequences. This observation opens for the possibility to generate non-natural functional RNA molecules. The idea has been explored recently in the case of proteins \cite{russ2020evolution}, its validation is forthcoming for RNA.
\end{enumerate}
Note that the applications to structure prediction are completely unsupervised: we use the parameters of a statistical model of an RNA family, but no direct structural data coming, e.g., from experimentally determined structures available in the Protein Data Bank\cite{wwpdb2019protein}. The quality of RNA structure prediction can be increased using supervised learning\cite{townshend2021geometric,zerihun2021coconet}, even if successes similar to proteins\cite{jumper2021highly} are still missing.

 \subsection{Direct-Coupling Analysis and Restricted Boltzmann Machines for Aptamer Design}
 
Statistical modeling from sequence  data was recently used  by some of us to reconstruct the binding fitness landscape of DNA aptamers \cite{DiGioacchino2022}. DNA aptamers are small   molecules with an hairpin structure (stem and loop), and even if they are DNA rather than RNA, the statistical modeling techniques are equivalent.
Laboratory selection experiments called SELEX\cite{tuerk1990systematic,ellington1990vitro} have been recently shown to be very powerful tools to design molecules with desired properties, such as clinical aptamers, mentioned in introduction, which avoid blood coagulation by strong binding to thrombin.
SELEX starts from an initial library of $10^{12}$ random initial  sequences for the two loops, each of 20 nucleotides, of a 2-loop aptamer. Sequences are  passed through 8 rounds of selection and amplification to extract very strongly bivalent binders to thrombin.  During the experiment $10^5$ sequences have been collected  starting from round 5 \cite{zhou2019dna}. 
Such  data have been used to infer the parameters of two generative models in machine learning, DCA as described above in Eq.~\eqref{eq:rna_dca} and Restricted Boltzmann Machines (RBM) described by the energy function 
\begin{equation}
 \label{eq:rna_rbm}
    H(r_1,...,r_L, y_1,...,y_P) = -\sum_{1\leq i \leq L} h_i(r_i) - \sum_{i,\mu} w_{i\mu} (r_i) \,y_i^{\mu}\ + \sum_{1\leq \mu \leq P} {\cal U}_{\mu}(y_i) \ .
\end{equation}
RBM are two-layer networks described by two sets of variables: (i) the variables $r_1,...,r_L$  in the visible layer, which stand for the RNA sequence, and (ii) the variables $y_1,...,y_P$  which act as feature detectors in the hidden layer. This simple bipartite network  can directly implement the dualism between data and their representations: extract features from data and generate data with desired features.  
RBM are also appealing in statistical mechanics because they are an extension of Hopfield models with  $P$ stored patterns \cite{barra2012equivalence}, for non binary patterns (Potts spins rather than Ising spins) and non quadratic hidden units. Using an appropriate regularisation, which implements sparsity conditions, the weights $w_{i\mu}$ have been shown to reflect key sequence motifs\cite{tubiana2019learning,shimagaki2019selection}: here the so called G-quadruplex motif  is important to  bind thrombin. 
The RBM and DCA models, learnt from the sequence data at round 6, are able to predict the most selected sequences at the next rounds: the log-likelihood of a sequence, corresponding to minus the RBM/DCA energy Eqs.~(\ref{eq:rna_rbm}) and~\eqref{eq:rna_dca}, can be used as a sequence score and is linearly correlated to the  fitness as measured by the sequence enrichment ratio (logarithm of the number of molecules selected at round $t+1$ divided the ones selected at the round $t$, i.e.~presented at round $t+1$). Moreover RBM can be used to design new binders which have not been found through the sequencing procedure, and to predict deleterious mutations which can interrupt binding \cite{DiGioacchino2022}.

 \section{RNA networks and the coordination of systemic regulatory programs}

\subsection{MicroRNAs and the ceRNA hypothesis}

As mentioned at the beginning of this chapter, it has long been known that not all RNA molecules are protein-coding \cite{cech2014noncoding}. For example, transfer RNAs (tRNAs), the best known among all non-coding RNAs (ncRNAs), play a key role in protein synthesis by carrying amino-acids to the elongation sites in ribosomes bound to protein-coding RNAs (messenger RNA, mRNA). In this process, the interaction between tRNA and mRNA is effectively catalyzed by another type of ncRNA, namely ribosomal RNA (rRNA). rRNA is especially remarkable because, besides being a key structural component of ribosomes, it constitutes over 80\% of the total RNA of animal cells\cite{zhao2018evaluation}. So most of our RNA is non-coding. In many organisms, then, stalled protein synthesis is rescued by another class of ncRNA known as transfer-messenger RNA (tmRNA), whose powers include the ability to degrade aberrant mRNA and to recycle the ribosome. We could continue with many more examples of ncRNAs from all domains of life acting in a myriad of biological processes. The list of known ncRNAs and their associated functions has indeed been growing steadily since the early 1980s, and with a particularly fast pace since the advent of high-throughput sequencing. But a whole new level of regulatory organisation managed by ncRNAs was uncovered when the mechanism known as RNA interference (RNAi) was first brought to light in 1998 \cite{fire1998potent}. 

In short, RNAi occurs when ncRNA molecules repress the expression of a gene by protein-mediated binding to specific sequence motifs on the mRNA. The most famous and possibly most important class of ncRNAs capable of performing RNAi in regulatory networks is formed by microRNAs (miRNAs\cite{gurtan2013role}), small (ca. 22 nt typically), highly conserved and highly stable eukaryotic RNA molecules whose life cycle can be summarized as follows\cite{jonas2015towards}: after being transcribed (from introns of protein-coding genes or miRNA-specific genes), they undergo a series of enzyme-catalyzed maturation steps before being incorporated into large protein complexes called RISCs (RNA-Induced Silencing Complexes); miRNA-RISCs expose specific nucleotide sequences through which they can bind short segments on their target RNAs called miRNA response elements (MREs); MREs are usually just 6 to 9 nt long,  implying relatively low binding affinities and therefore a potentially significant interference from other short sequences to which they can bind in purely electrostatic manner; miRNAs can target both coding and non-coding RNAs; if the target is a mRNA, the bulky RISC bound to the mRNA impedes translation by blocking the ribosome; finally, miRNA-target complex degradation is elicited and can occur through two distinct pathways: a so-called `catalytic' pathway leading to the re-cycling of the miRNA; and a `stoichiometric' pathway leading to the degradation of both miRNA and target\cite{baccarini2011kinetic}. 

If one focuses on sequence-specific interactions, different types of miRNAs have different binding sites and hence repress different genes, often acting within clearly identifiable regulatory motifs whose functions have been subject of much work\cite{re2017microrna}. On a broader scale, however, the global interaction network linking miRNAs to their targets (both coding and non-coding) spans across the entire transcriptome and is strongly heterogeneous, as each miRNA can repress many distinct RNA targets and, vice-versa, each RNA species can be the target of many miRNA species\cite{franks2017post}. Cells therefore invest serious molecular and energy resources (e.g.~via specialized maturation enzymes, RISC proteins, biosynthetic machinery, etc.) into maintaining and conserving across generations functional batteries of miRNAs together with their interaction networks, despite the fact that the individual miRNA-target coupling, while evolutionarily preserved, can be rather weak. Given that gene expression can be repressed in many ways, the obvious question is what makes miRNAs so convenient for cells.

Two answers have so far attracted the largest attention. In the first scenario, miRNAs provide an essential contribution to the fine tuning of gene expression levels\cite{baek2008impact}. It is well known that protein levels are tightly regulated in higher organisms. This requires molecular mechanism that are capable of processing the inherent stochasticity of regulatory processes (e.g.~due to transcription factors diffusing to their binding sites, bursty transcription events, etc). Experimental and theoretical work has indeed shown that gene expression noise is reduced in presence of miRNAs, albeit not always by spectacular amounts compared to the unregulated case or to alternative regulatory means\cite{siciliano2013mirnas,schmiedel2015microrna,riba2014combination}. Besides their ability to repress gene expression, this noise-reducing capability is without doubt an added value of miRNAs.

\begin{figure}[t]
\begin{center}
\includegraphics[width=1.0\columnwidth]{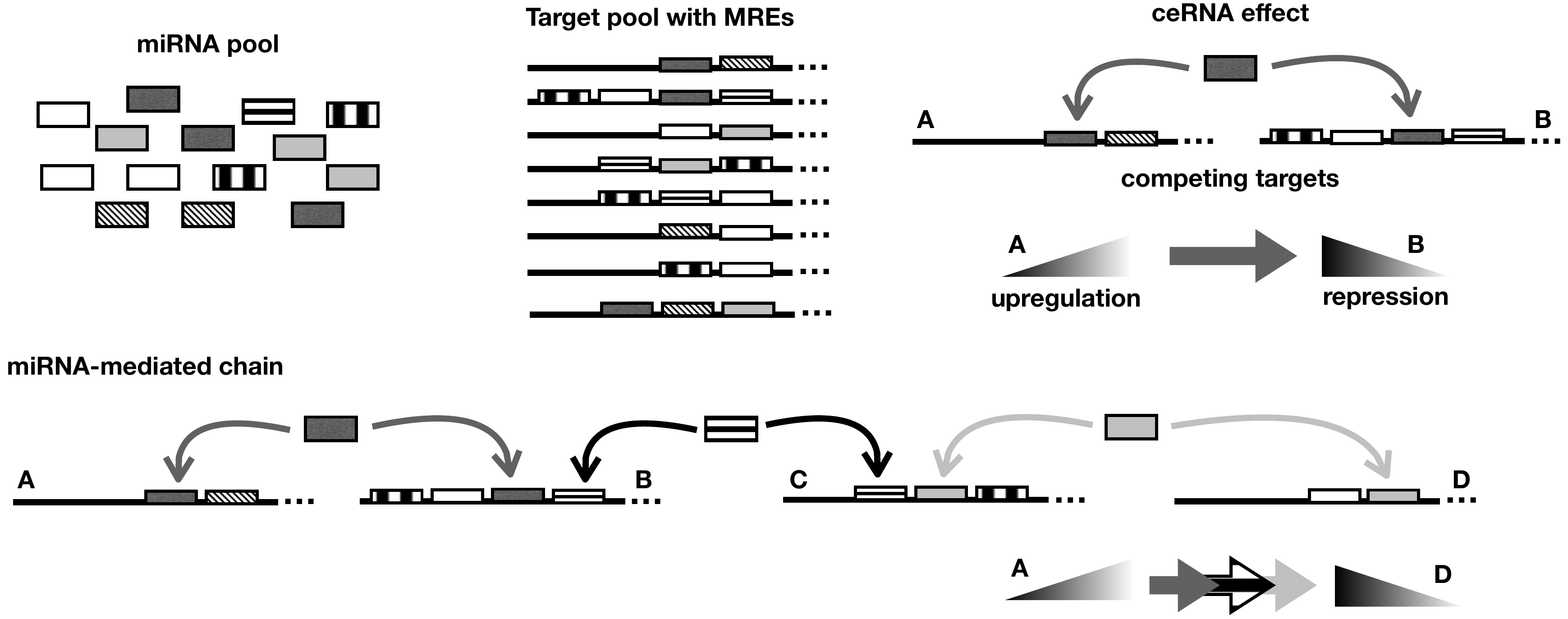}
\caption{\label{mirnanet}A diverse pool of miRNAs coexists in cells with a diverse pool of targets. Targets sharing at least one miRNA species as regulator compete to bind it. If the level of target A is increased while keeping all else fixed, the likelihood of the miRNA binding target A increases. In turn, repression is increasingly lifted off target B, which could then become more likely available e.g. for translation. This effect can propagate between targets that do not share any common miRNA regulators, such as A and D above, through chains of miRNA-mediated couplings induced by the same competition mechanism.}
\end{center}
\end{figure}

The second scenario adds a totally different ingredient, as it postulates that miRNAs can establish weak but extended crosstalk interaction networks between their targets, which effectively allow for a degree of systemic cross-modulation of expression levels\cite{salmena2011cerna}. In other words, miRNAs orchestrate a weak but large-scale regulatory network that can (weakly) coordinate the translation levels of {\it all} RNAs in the transcriptome. The key to unlock this scenario is competition: if a miRNA species interacts with two different targets (say $A$ and $B$), an increase in the availability of $A$ will draw more miRNAs towards it and therefore lift repression, at least partially, from $B$. In other words, upon increasing the expression level of one miRNA target, the expression level of the other increases as well (Fig.~\ref{mirnanet}). 
It is easy to see how such an effect can effectively link the entire transcriptome via chains of miRNA-target interactions\cite{nitzan2014interactions}. Unfortunately, quantifying its relevance in a physiological context is hard due to the weakness of the individual miRNA-target interactions\cite{denzler2014assessing,bosson2014endogenous}, although traces of its action can be found in indirect effects\cite{martirosyan2017cerna}. The main questions are the following: under which circumstances can the myriad of weak miRNA-target couplings encoded in the RNA sequences build up to generate coherent target-target interactions spanning across the whole transcriptome? Can these interactions affect system-level outcomes? Are these properties embedded in the specific topology of the miRNA-target network? How strongly do they depend on the interaction parameters (e.g. the binding affinities)? 

In view of the experimental limitations, some understanding can only be achieved through a theoretical approach. Luckily, all of these questions echo classical problems in statistical mechanics and complexity theory. Methods developed in these fields (and, as we shall see, for disordered systems specifically) are therefore ideally suited for this task.

\subsection{A statistical theory of competition in RNA networks}

A (simplified) theory of RNA networks starts from the dynamics of molecular levels. Denoting respectively by $\mu_a$ ($a=1,\ldots,M$) and $m_i$ ($i=1,\ldots,N$) the levels of miRNA species $a$ and target species $i$, and using the shorthand $\dot{x}\equiv\frac{dx}{dt}$, the deterministic mass action kinetics of a miRNA-target network is described by (see e.g. \cite{figliuzzi2013micrornas,bosia2013modelling,noorbakhsh2013intrinsic})
\begin{gather}
\dot \mu_a=\beta_a-\delta_a \mu_a-\sum_i k_{ia}^+ m_i\mu_a + \sum_{i}(k_{ia}^-+\kappa_{ia})c_{ia}~~,\label{murna}\\
\dot m_i=b_i-d_i m_i-\sum_a k_{ia}^+ m_i\mu_a +\sum_a k_{ia}^- c_{ia}~~, \label{uno}\\
\dot c_{ia}= k_{ia}^+ m_i \mu_a-(\sigma_{ia}+\kappa_{ia}+k_{ia}^-)c_{ia}~~,\label{complex}
\end{gather}
where $c_{ia}$ stands for the level of miRNA-target complexes while synthesis rates ($b_i,\beta_a$), degradation rates ($d_i$,$\delta_a$), association and dissociation rates ($k_{ia}^{\pm}$) and complex processing rates ($\sigma_{ia},\kappa_{ia}$) represent kinetic parameters whose values are mostly unknown (and likely unknowable). To get some intuition of what this system looks like if seen through a statistical-physics lens, one can first observe that, if $\sigma_{ia}+\kappa_{ia}+k_{ia}^-$ is much larger than both $d_i$ and $\delta_a$, Eq.~(\ref{complex}) equilibrates much faster than the other two. In this limit, and under reasonable assumptions for kinetic rates, the steady state values of $m_i$ and $\mu_a$ can be found by minimizing the (Lyapunov) function \cite{martirosyan2017translating}
\begin{gather}\label{elle}
L(\{m_i\},\{\mu_a\})\!=\!\sum_{i,a}k_{ia}^+ m_i\mu_a\!-\!\sum_i(b_i\log m_i\!-\!d_im_i)\!-\!\sum_a(\beta_a\log \mu_a\!-\!\delta_a \mu_a).
\end{gather}
It is easy to see that, for any physically consistent choice of the kinetic parameters, $L$ has a unique non-trivial minimum (i.e.~with non-zero concentration vectors). To a first approximation, then, stochastic effects that were neglected in writing Eqs.~(\ref{murna}-\ref{complex}) can be captured by assuming that fluctuations around the minimum of $L$ correspond to sampling molecular levels from a Boltzmann-Gibbs distribution
\begin{gather}\label{gibbs}
P(\{m_i\},\{\mu_a\})=\frac{e^{-L/T}}{Z(T)}~~~~~,~~~~~ Z(T)=\sum_{\{m_i\},\{\mu_a\}}\, e^{-L/T}~~.
\end{gather}
where $T>0$ is a (fictitious) ``temperature'' parameter that is expected to correlate with the strength of the molecular noise to be added to  Eqs.~(\ref{murna}-\ref{complex}) for a full stochastic description. The above considerations suggest a theoretical approach based on the following points: 
\begin{enumerate}
\item[(i)] The {\it intensity} of miRNA-mediated crosstalk between targets $i$ and $j$ at equilibrium (i.e. at stationarity) can be quantified via the susceptibility\cite{figliuzzi2013micrornas}
\begin{gather}
\chi_{ij}=\frac{\partial \langle{m_i}\rangle}{\partial b_j}~~,
\end{gather}
where $\langle{\cdots}\rangle$ denotes equilibrium values. $\chi_{ij}$ can be computed either from the steady states of the dynamics or within a purely statistical mechanics framework in which $\langle{\cdots}\rangle=\sum_{\{m_i\},\{\mu_a\}}\,\cdots P(\{m_i\},\{\mu_a\})$ (see Eq.~(\ref{gibbs})). Importantly, because $\langle{m_i}\rangle=-T\frac{\partial}{\partial d_i}\log Z(T)$, one can immediately see that $\chi_{ij}\neq\chi_{ji}$: miRNA-mediated crosstalk interactions are non-symmetric.
\item[(ii)] The cross-regulatory equilibrium {\it effectiveness} of miRNA-mediated interactions versus the standard regulatory control exerted on a target (say $j$) by its transcription factor (TF) can be quantified by comparing the mutual informations $I(m_j,f_j)$ and $I(m_j,f_i)$, where $f_j$ (resp. $f_i$) denotes the level of the transcription factor controlling $j$ (resp. one of $j$'s competitors to bind miRNAs), while as usual
\begin{gather}
I(m_j,f)=\int df\, P(f)\int dm_j\,P(m_j|f)\log_2\frac{P(m_j|f)}{P(m_j)}~~.
\end{gather}
One would specifically like to identify the conditions under which miRNA-mediated control and TF-mediated control are equally effective. Such an analysis only requires a slight generalization of Eqs.~(\ref{murna})-(\ref{complex}) to include transcription factors. In the above formula, $P(m_j)$ and $P(m_j|f)$ represent, respectively, the equilibrium distributions of the level of target $j$ and of the level of target $j$ given the level of TF $f$, whereas $P(f)$ stands for the probability density of transcription factor levels. The latter quantity is treated as an exogenous variable. In fact, $I$ depends on the choice of $P(f)$. To establish a benchmark, it is therefore useful to compare the optimal values of $I(m_j,f_j)$ and $I(m_j,f_i)$ obtained by maximizing the two quantities with respect to $P(f_j)$ and $P(f_i)$, respectively.\cite{tkavcik2008information,martirosyan2016probing}
\item[(iii)] As the exact values of kinetic parameters are  impossible to quantify in practice even for small systems, the {\it typical (or context-independent) large-scale properties} of the emerging cross-talk network at stationarity can be found by analysing, for large $N$ and $M$, averages of physical observables like expression levels or susceptibilities over ensembles of kinetic parameter values\cite{miotto2019competing}. In such a scenario, kinetic parameters are treated as {\it quenched disorder}, so that the function $L$ in Eq.~(\ref{elle}) effectively becomes a {\it disordered cost function} whose minima depend on the specific realization of the disorder. The {\it stability} of expression profiles and their {\it robustness} against changes in kinetic parameters would immediately follow from this analysis. Notice that typical properties are physiologically relevant, as they are expected to hold (for a large network) independently of the specific values of the parameters.
\item[(iv)] {\it Off-equilibrium properties}, describing e.g. the approach to equilibrium or the transient response of the network to perturbations, can instead be derived by studying the linearized version of Eqs.~(\ref{murna}-\ref{complex}) in the limit of small perturbations.\cite{figliuzzi2014rna}
\end{enumerate}
The above program has been carried out starting in 2013 over a series of papers by various authors, covering aspects ranging from the off-equilibrium dynamics of small miRNA-mediated circuits to the typical system-level properties of crosstalk in the human transcriptome (see Ref.~\cite{martirosyan2019kinetic} for a thorough review). For sakes of clarity, we focus here on two sets of results with high biological significance, concerning the roles of (a) transcriptional heterogeneities and (b) topological heterogeneities in shaping the emergent network-scale expression profiles in the human miRNA interactome mapped via the CLASH protocol (Crosslinking, Ligation And Sequencing of Hybrids), accounting for $\mathcal{O}(10^7)$ potential miRNA-mediated cross-regulatory interactions, each quantifiable by the value of $\chi_{ij}$.
\begin{description}
\item[Transcriptional heterogeneities] -- Estimated transcription rates vary drastically from one transcript to another, as some genes are naturally much more expressed than others. An important question is how crosstalk patterns, i.e.~the statistics of $\boldsymbol{\chi}=\{\chi_{ij}\}$ that is generated between miRNA targets in an extended network of miRNA-target interactions, are affected by such variability. Results based on averaging over ensembles of transcription rates have shown that \cite{miotto2019competing}:
\begin{itemize}
\item[-] The average intensity of crosstalk interactions, namely $\overline{\langle\chi\rangle}$ (with the inner average over pairs of targets and the outer average over disorder) is weakly dependent on the strength of transcriptional heterogeneities: the typical susceptibility of the network is robust to baseline gene expression variability. 
\item[-] On the other hand, the maximum achievable value of $\chi_{ij}$ increases as transcription rates become more heterogeneous: in other words, heterogeneous kinetic parameters favour the establishment of stronger crosstalk links. Importantly, the stability of expression profiles is maximal (i.e. sample-to-sample fluctuations are smaller) when the strongest crosstalk is achieved. 
\item[-] In view of this, crosstalk gets more {\it selective} as heterogeneity increases, i.e., in each condition, only a sub-network of cross-regulatory couplings is effectively active. 
\item[-] Perhaps most remarkably, heterogeneities enhance crosstalk non-locality, as crosstalk patterns become less and less correlated with local interaction parameters (e.g. association rates) as transcriptional variability increases. In practice, extended chains of strong miRNA-mediated crosstalk interactions become more and more frequent as heterogeneity increases. In this respect, one can say that the highly promiscuous miRNAs exploit the large-scale `disorder' of kinetic parameters to implement a weak but system-level regulatory layer that coordinates and stabilizes expression levels. 
\end{itemize} 
\item[Topological heterogeneities] -- Evolution in the miRNA interactome effectively acts on the structure of miRNA-target interactions. To what degree are crosstalk patterns controlled by the specific wiring of the miRNA-target network? To answer this question, one can re-analyze the emergent crosstalk patterns after properly randomizing the empirical interactome. The surprising results\cite{miotto2019competing} is that all properties discussed above survive as long as the randomization protocol preserves the degrees of the nodes (i.e. the number of in-coming and out-going links). By contrast, any randomization that alters the degree sequence (for instance by making the network more homogeneous through the re-wiring of the most connected species) generates weaker and more local crosstalk interactions, while the stability of expression profiles is higher in absence of crosstalk. This strongly suggests that certain features of miRNA-mediated RNA crosstalk patterns (selectivity, maximal intensity, stability etc.) are encoded in the detailed structure of the miRNA-target interaction network by natural selection.
\end{description}
System-level regulatory mechanisms are likely ubiquitous (especially if driven by competition for limited molecular resources) and have just begun to be understood. There is little doubt in our view that the concepts and tools developed for spin glasses and complex physical systems will play a major role in the analysis of these phenomena (and of their physiological relevance) in the coming years.

\section*{Acknowledgments}

We are deeply indebted to Giorgio Parisi for many years -- past and future -- of inspiring discussions and collaborations. We are likewise grateful to the many colleagues and friends with whom some of the ideas presented in this chapter have been developed. ADM, AP, and MW wish to acknowledge partial financial support from the EC funded Marie Skłodowska-Curie program under grant agreement no. 734439 (InferNet).

\bibliographystyle{ws-rv-van}
\bibliography{chapter_bio}

\end{document}